\documentclass[pdflatex,sn-mathphys]{sn-jnl}

\usepackage{caption, booktabs}
\usepackage{graphicx}
\usepackage{etoolbox}
    \makeatletter
    \patchcmd{\ps@headings}
    {\hbox to \hsize{\hfill Springer Nature 2021 \LaTeX\ template\hfill}}
    {\hbox to \hsize{}}
    {}
    {}
    \patchcmd{\ps@headings}
    {\hbox to \hsize{\hfill Springer Nature 2021 \LaTeX\ template\hfill}}
    {\hbox to \hsize{}}
    {}
    {}
    \patchcmd{\ps@titlepage}
    {\hbox to \hsize{\hfill Springer Nature 2021 \LaTeX\ template\hfill}}
    {\hbox to \hsize{}}
    {}
    {}
    \makeatother

\jyear{2021}%

\theoremstyle{thmstyleone}%
\theoremstyle{thmstyletwo}%

\theoremstyle{thmstylethree}%

\begin{document}

\title {Fluorescence angiography classification in colorectal surgery}
\subtitle{A preliminary report}

\author*[1,2]{\fnm{Antonio} \sur{S Soares}}\email{antonio.soares.17@ucl.ac.uk}

\author[1,3]{\fnm{Sophia} \sur{Bano}} \email{sophia.bano@ucl.ac.uk}

\author[1,4]{\fnm{Neil} \sur{T Clancy}} \email{n.clancy@ucl.ac.uk}

\author[1,2]{\fnm{Laurence} \sur{B Lovat}} \email{l.lovat@ucl.ac.uk}

\author[1,2]{\fnm{Danail} \sur{Stoyanov}} \email{danail.stoyanov@ucl.ac.uk}

\author[1,2]{\fnm{Manish} \sur{Chand}} \email{m.chand@ucl.ac.uk}

\affil*[1]{\orgdiv{Wellcome/EPSRC Interventional and Surgical Sciences (WEISS) Centre}, \orgname{University College London}, \orgaddress{\street{43-45 Foley Street}, \city{London}, \postcode{W1W 7JN}, \country{United Kingdom}}}

\affil[2]{\orgdiv{Division of Surgery and Interventional Sciences}, \orgname{University College London}}

\affil[3]{\orgdiv{Department of Computer Science}, \orgname{University College London}}

\affil[4]{\orgdiv{Department of Medical Physics and Biomedical Engineering}, \orgname{University College London}}

\maketitle
\newpage

\section{Abstract}\label{sec1}

\begin{abstract}

\textbf{Background:} Fluorescence angiography has shown very promising results in reducing anastomotic leaks by allowing the surgeon to select optimally perfused tissue. However, subjective interpretation of the fluorescent signal still hinders broad application of the technique, as significant variation between different surgeons exists. Our aim is to develop an artificial intelligence algorithm to classify colonic tissue as 'perfused' or 'not perfused' based on intraoperative fluorescence angiography data. 
 
\textbf{Methods:} A classification model with a Resnet architecture was trained on a dataset of fluorescence angiography videos of colorectal resections at a tertiary referral centre. Frames corresponding to fluorescent and non-fluorescent segments of colon were used to train a classification algorithm. Validation using frames from patients not used in the training set was performed, including both data collected using the same equipment and data collected using a different camera. Performance metrics were calculated, and saliency maps used to further analyse the output. A decision boundary was identified based on the tissue classification.  

\textbf{Results:} A convolutional neural network was successfully trained on 1790 frames from 7 patients and validated in 24 frames from 14 patients. The accuracy on the training set was 100\%, on the validation set was 80\%. Recall and precision were respectively 100\% and 100\% on the training set and 68.8\% and 91.7\% on the validation set. 
 
\textbf{Conclusion:} Automated classification of intraoperative fluorescence angiography with a high degree of accuracy is possible and allows automated decision boundary identification. This will enable surgeons to standardise the technique of fluorescence angiography. A web based app was made available to deploy the algorithm.

\end{abstract}




\newpage

\section{Introduction}\label{sec2}
Anastomotic leak is an important source of morbidity, mortality and rising healthcare costs \cite{Ashraf2013} in colorectal surgery. Surgeons use fluorescence angiography to optimise the perfusion of the tissue used to create the anastomosis\cite{Keller_ICG}. The use of fluorescence angiography has been shown to reduce anastomotic leaks in non-randomized studies \cite{Blanco-Colino2018, Degett2016} and has been recommended by the European Association for Endoscopic Surgery [submitted]. The true utility of the technique remains unquantified.   

The interpretation of the signal in fluorescence angiography is dependent on the surgeon and perhaps represents the actual ‘learning curve’ for this technique. These decisions have a high degree of variability shown both for static images [Soares, AS et al., Interobserver agreement study, \textit{under review}] and dynamic video \cite{hardy2021inter} assessments. Automating the interpretation of the signal would eliminate this variability and provide an alternative means of quantification, thereby leveraging the usefulness of surgical data science \cite{Maier-Hein2022} in the clinical setting.

Convolutional neural networks are especially suited for image analysis. The input for these neural networks are the pixel values of a given image and the outputs are the probabilities of the image belonging to a particular pre-defined class. 

This paper describes the development of an artificial intelligence algorithm to automatically classify tissue based on perfusion using fluorescence angiography during colorectal surgery. A decision boundary is estimated based on the classification of different areas of the tissue of interest.

\section{Methods}\label{sec3}

\subsection{Dataset}\label{subsec1}
Patients undergoing colorectal surgery that involved construction of an anastomosis were included in the study. De-identified data was collected between May and November 2019 as previously described by our group \cite{soares2022multisensor}. Additional data used for validation was collected between January and August 2021. This was a convenience sample.
In all cases, the bowel transection was performed following extracorporealisation of the bowel. Prior to transection, the bowel was placed on clean white swabs to act as ‘background’. A continuous sequence was recorded from the point of ICG injection. 10 mg of ICG was injected intravenously in all cases. The recording lasted until the operating surgeon decided fluorescence intensity was adequate to choose the point of proximal transection (maximal fluorescence intensity), as per current standard practice in FA. The training data was collected using the PINPOINT camera (Stryker, Michigan, USA) and the validation data was collected with PINPOINT and Stryker 1688 (Stryker, Michigan, USA) and Arthrex (Arthrex Inc., Florida, USA) cameras.
All patients provided written informed consent for video collection, and standard clinical practice was followed in all cases. No missing data occurred. The TRIPOD statement\cite{TRIPOD} was followed.

\subsection{Model development}\label{subsec2}
In the literature of deep learning, convolutional neural networks (CNNs) have shown remarkable performance for the classification task. Transfer learning was used based on a model trained on ResNet34. The neural network was then fine-tuned using the Fast AI library in Python \cite{howard2018fastai}. Before loading the data for training, it was augmented using a randomly selected crop from the complete 1440 x 1080 pixel frame. This process occurred for 4 epochs. A random selection of 20\% of the training data was used for initial validation. The threshold defined for classification of fluorescent was a probability output of the algorithm above 80\%. The final result is an algorithm which takes the colorectal image as input and output a binary decision, whether the given input image has fluorescence in it or not.

\subsection{Model evaluation}\label{subsec3}
The model was deployed through a web interface using the Gradio Library and Huggingface Spaces (\url{https://huggingface.co/spaces/asampaiosoares/fluorescence_id_app}). This platform enabled the serial testing of additional frames that were not used for training the algorithm. Saliency maps were created to analyse the predictions from the validation set.

\subsection{Decision boundary estimation}\label{subsec4}
The estimation of the decision boundary was performed on the images considered positive for fluorescence in the validation set. These images were divided into JPEG segments of 100 pixel width along the colonic longitudinal axis using imageJ \cite{Schneider2012} and inputted into the algorithm. Both the binary classification results as well as the relative probabilities were recorded for each case. The decision boundary was defined as the limit between the most distal area that was classified as fluorescent and the areas considered not fluorescent.

\subsection{Model deployment}\label{subsec5}
The algorithm was deployed through an online platform called Gradio, which allows the upload of images to be classified using the algorithm developed. 

\section{Results}\label{sec4}
\begin{figure}[H]
\centering
\includegraphics[scale=0.3]{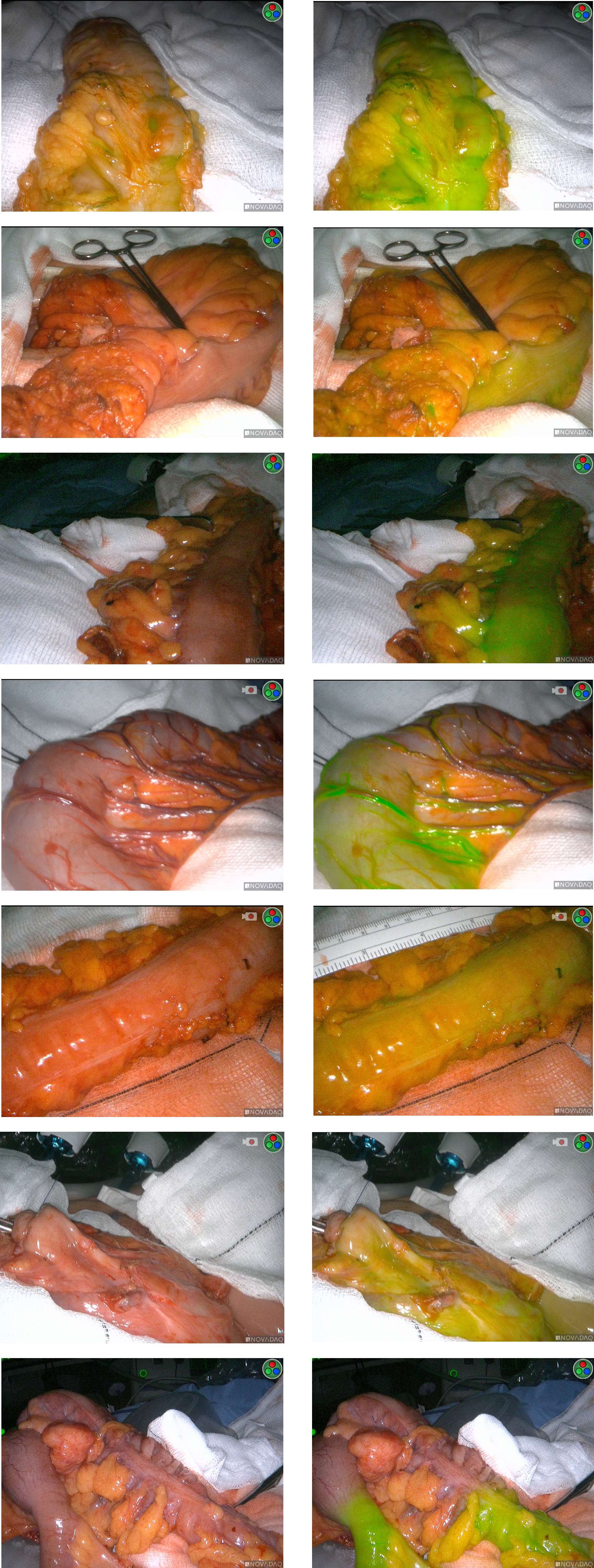}
\caption{Representative frames of the training set}
\label{fig:training_frames}
\end{figure}

A dataset of patients undergoing colorectal resection was created with the characteristics shown in Table \ref{my-table1}. Representative frames corresponding to the training set can be found in figure \ref{fig:training_frames}. 

\begin{table}[ht]
\centering
\begin{tabular}{@{}lcccc@{}}
\toprule
                             & \textbf{Training set} & \textbf{Validation set} & \textbf{} & \textbf{} \\ \midrule
Number of patients           & 7                    & 14                      &           &           \\
Number of frames             & 1790                 & 30                      &           &           \\
Fluorescence positive frames & 19.6\%               & 53.3\%                  &           &           \\
\multicolumn{1}{c}{}         &                      &                         &           &           \\ \bottomrule
\end{tabular}
\caption{Dataset characteristics}
\label{my-table1}
\end{table}

\subsection{Training and validation}\label{subsec1}

Training took 21 minutes and 11 seconds. The validation on data not used in the training set was done manually using the Gradio app. The performance metrics can be seen in Table \ref{my-table2}.

\begin{figure}[H]
\centering
\includegraphics[scale=0.5]{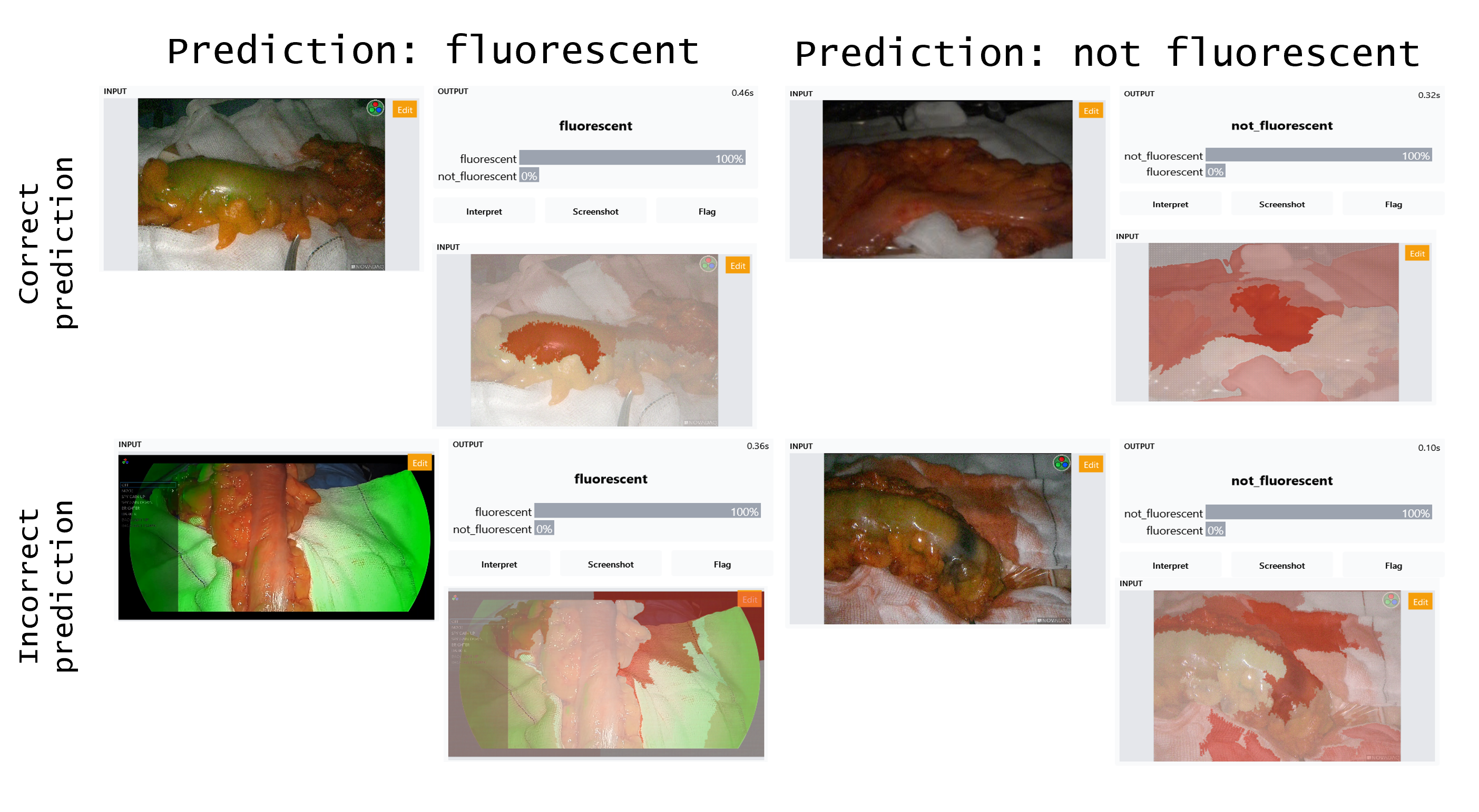}
\caption{Representative frames of predictions performed by the algorithm}
\label{fig:rep_frames}
\end{figure}

\begin{table}[ht]
\centering
\resizebox{1.0\textwidth}{!}{
\begin{tabular}{@{}ccccc@{}}
\toprule
 & \textbf{Recall / Sensitiviy} & \textbf{Precision / PPV} & \textbf{Accuracy} & \textbf{F1 score} \\ \midrule
Training                   & 100.0\% & 100.0\% & 100.0\% & 100.0\% \\
Validation overall         & 68.8\%  & 91.7\%  & 80.0\%  & 78.6\%  \\
Validation - internal data & 60.0\%  & 100.0\% & 80.0\%  & 75.0\%  \\
Validation - external data & 83.3\%  & 83.3\%  & 80.0\%  & 83.3\%  \\ \bottomrule
\end{tabular}
}
\caption{Performance metrics for the classification algorithm. PPV positive predictive value}
\label{my-table2}
\end{table}
Representative frames of true positive, true negative, false positive and false negative results can be seen in figure \ref{fig:rep_frames}, including the respective saliency maps. These maps show areas that contribute more to activation of the neural network in a darker shade of red. 

\subsection{Decision boundary estimation}\label{subsec2}
\begin{figure}[H]
\centering
\includegraphics[scale=0.3]{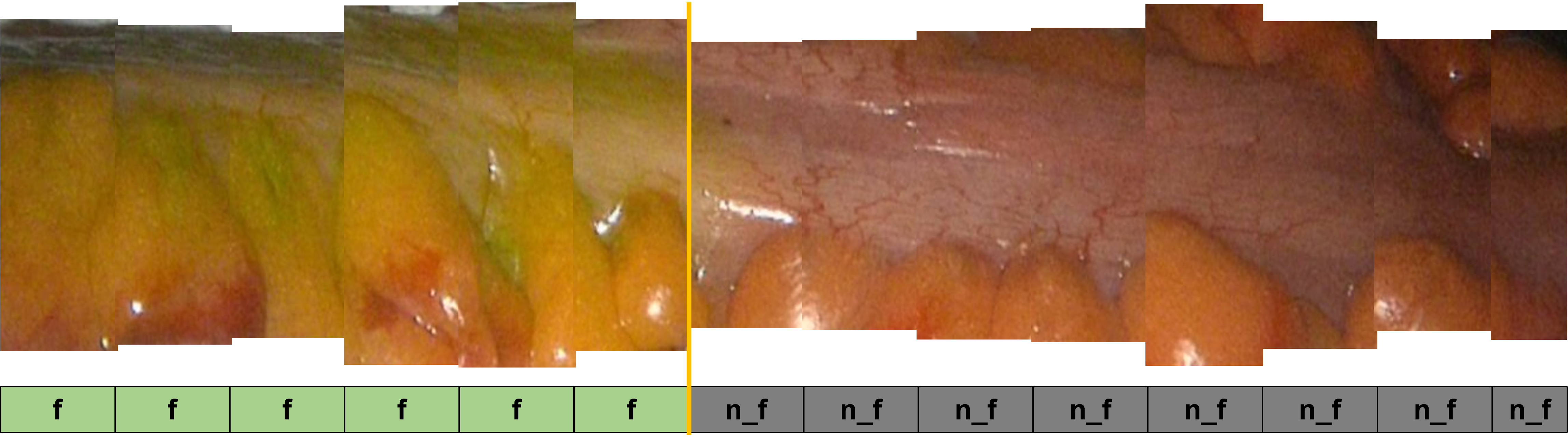}
\caption{Representative frame for decision boundary analysis, where green boxes denote segments of the image classified as fluorescent and grey boxes denote segments classified as not fluorescent with the decision boundary between the two, shown in yellow}
\label{fig:representative_frame_decision_boundary}
\end{figure}

The decision boundary was estimated for all the positive cases used for validation as shown in figure \ref{fig:representative_frame_decision_boundary}.  

\section{Discussion}\label{sec5}
These results have show that a neural network trained on clinical data has achieved an accuracy over 90\%. There was a drop in performance using data acquired with a different equipment for validation but still maintaining an accuracy over 80\%. 

The analysis of the saliency maps shows that in the cases where both false negative and positive results were obtained the pixels corresponding to the colon were not responsible for the main activation of the algorithm. This leads us to hypothesize that an image analysis pipeline segmenting the colon from the background first and, on this output, running the classification algorithm could lead to improved results. Based on previous work from our group [Soares, AS et al., Interobserver agreement study, \textit{under review}] and others \cite{hardy2021inter} significant variation exists between different surgeons' assessments. Standardisation of fluorescence angiography is a necessary step to enable accurate comparison of results. 

Our algorithm provides an easily accessible tool based on state of the art deep learning techniques that eliminates interobserver variability in interpreting the fluorescent signal. Previous efforts to standardise fluorescence angiography assessments have focused on quantitative measurements requiring post hoc signal processing \cite{wada2017icg, son2019quantitative}. Our algorithm has the ability to simplify post-hoc processing as it was trained on data captured intraoperatively and is designed to output a binary output ("fluorescent" vs "non fluorescent").

Some limitations should be addressed. Although a high number of frames was used for training, the variability of frames used for validation was high but the overall number was small. This should be mitigated by including data from more centres performing fluorescence angiography. Furthermore, the data collected was of low quality to begin with, which limited the usefulness of the algorithm. This should be interpreted with caution as even an experienced surgeon would find some trouble analysing this data intraoperatively. 
A close collaboration between engineers and clinicians has led to the development of the algorithm presented in this work. Further work should aim to enrich the dataset for further refinement of the algorithm. Clinical testing in a randomized clinical trial will be necessary before large scale implementation can occur.  

\section{Conclusion}\label{sec6}
An artificial intelligence algorithm to classify colonic perfusion based on fluorescence angiography was developed using state of the art deep learning techniques. This work paves the way for standardisation of fluorescence angiography in colorectal surgery.

\bibliography{sn-bibliography}


\begin{thebibliography}{12}
\ifx \bisbn   \undefined \def \bisbn  #1{ISBN #1}\fi
\ifx \binits  \undefined \def \binits#1{#1}\fi
\ifx \bauthor  \undefined \def \bauthor#1{#1}\fi
\ifx \batitle  \undefined \def \batitle#1{#1}\fi
\ifx \bjtitle  \undefined \def \bjtitle#1{#1}\fi
\ifx \bvolume  \undefined \def \bvolume#1{\textbf{#1}}\fi
\ifx \byear  \undefined \def \byear#1{#1}\fi
\ifx \bissue  \undefined \def \bissue#1{#1}\fi
\ifx \bfpage  \undefined \def \bfpage#1{#1}\fi
\ifx \blpage  \undefined \def \blpage #1{#1}\fi
\ifx \burl  \undefined \def \burl#1{\textsf{#1}}\fi
\ifx \doiurl  \undefined \def \doiurl#1{\url{https://doi.org/#1}}\fi
\ifx \betal  \undefined \def \betal{\textit{et al.}}\fi
\ifx \binstitute  \undefined \def \binstitute#1{#1}\fi
\ifx \binstitutionaled  \undefined \def \binstitutionaled#1{#1}\fi
\ifx \bctitle  \undefined \def \bctitle#1{#1}\fi
\ifx \beditor  \undefined \def \beditor#1{#1}\fi
\ifx \bpublisher  \undefined \def \bpublisher#1{#1}\fi
\ifx \bbtitle  \undefined \def \bbtitle#1{#1}\fi
\ifx \bedition  \undefined \def \bedition#1{#1}\fi
\ifx \bseriesno  \undefined \def \bseriesno#1{#1}\fi
\ifx \blocation  \undefined \def \blocation#1{#1}\fi
\ifx \bsertitle  \undefined \def \bsertitle#1{#1}\fi
\ifx \bsnm \undefined \def \bsnm#1{#1}\fi
\ifx \bsuffix \undefined \def \bsuffix#1{#1}\fi
\ifx \bparticle \undefined \def \bparticle#1{#1}\fi
\ifx \barticle \undefined \def \barticle#1{#1}\fi
\bibcommenthead
\ifx \bconfdate \undefined \def \bconfdate #1{#1}\fi
\ifx \botherref \undefined \def \botherref #1{#1}\fi
\ifx \url \undefined \def \url#1{\textsf{#1}}\fi
\ifx \bchapter \undefined \def \bchapter#1{#1}\fi
\ifx \bbook \undefined \def \bbook#1{#1}\fi
\ifx \bcomment \undefined \def \bcomment#1{#1}\fi
\ifx \oauthor \undefined \def \oauthor#1{#1}\fi
\ifx \citeauthoryear \undefined \def \citeauthoryear#1{#1}\fi
\ifx \endbibitem  \undefined \def \endbibitem {}\fi
\ifx \bconflocation  \undefined \def \bconflocation#1{#1}\fi
\ifx \arxivurl  \undefined \def \arxivurl#1{\textsf{#1}}\fi
\csname PreBibitemsHook\endcsname

\bibitem{Ashraf2013}
\begin{barticle}
\bauthor{\bsnm{Ashraf}, \binits{S.Q.}},
\bauthor{\bsnm{Burns}, \binits{E.M.}},
\bauthor{\bsnm{Jani}, \binits{A.}},
\bauthor{\bsnm{Altman}, \binits{S.}},
\bauthor{\bsnm{Young}, \binits{J.D.}},
\bauthor{\bsnm{Cunningham}, \binits{C.}},
\bauthor{\bsnm{Faiz}, \binits{O.}},
\bauthor{\bsnm{Mortensen}, \binits{N.J.}}:
\batitle{{The economic impact of anastomotic leakage after anterior resections
  in English NHS hospitals: Are we adequately remunerating them?}}
\bjtitle{Colorectal Disease}
\bvolume{15}(\bissue{4}),
\bfpage{190}--\blpage{199}
(\byear{2013}).
\doiurl{10.1111/codi.12125}
\end{barticle}
\endbibitem

\bibitem{Keller_ICG}
\begin{barticle}
\bauthor{\bsnm{Keller}, \binits{D.S.}},
\bauthor{\bsnm{Ishizawa}, \binits{T.}},
\bauthor{\bsnm{Cohen}, \binits{R.}},
\bauthor{\bsnm{Chand}, \binits{M.}}:
\batitle{Indocyanine green fluorescence imaging in colorectal surgery:
  overview, applications, and future directions}.
\bjtitle{The Lancet Gastroenterology 'I\&' Hepatology}
\bvolume{2}(\bissue{10}),
\bfpage{757}--\blpage{766}
(\byear{2017}).
\doiurl{10.1016/S2468-1253(17)30216-9}
\end{barticle}
\endbibitem

\bibitem{Blanco-Colino2018}
\begin{barticle}
\bauthor{\bsnm{Blanco-Colino}, \binits{R.}},
\bauthor{\bsnm{Espin-Basany}, \binits{E.}}:
\batitle{{Intraoperative use of ICG fluorescence imaging to reduce the risk of
  anastomotic leakage in colorectal surgery: a systematic review and
  meta-analysis.}}
\bjtitle{Techniques in coloproctology}
\bvolume{22}(\bissue{1}),
\bfpage{15}--\blpage{23}
(\byear{2018}).
\doiurl{10.1007/s10151-017-1731-8}
\end{barticle}
\endbibitem

\bibitem{Degett2016}
\begin{barticle}
\bauthor{\bsnm{Degett}, \binits{T.H.}},
\bauthor{\bsnm{Andersen}, \binits{H.S.}},
\bauthor{\bsnm{G{\"{o}}genur}, \binits{I.}}:
\batitle{{Indocyanine green fluorescence angiography for intraoperative
  assessment of gastrointestinal anastomotic perfusion: a systematic review of
  clinical trials}}.
\bjtitle{Langenbeck's Archives of Surgery}
\bvolume{401}(\bissue{6}),
\bfpage{767}--\blpage{775}
(\byear{2016}).
\doiurl{10.1007/s00423-016-1400-9}
\end{barticle}
\endbibitem

\bibitem{hardy2021inter}
\begin{botherref}
\oauthor{\bsnm{Hardy}, \binits{N.P.}},
\oauthor{\bsnm{Dalli}, \binits{J.}},
\oauthor{\bsnm{Khan}, \binits{M.F.}},
\oauthor{\bsnm{Andrejevic}, \binits{P.}},
\oauthor{\bsnm{Neary}, \binits{P.M.}},
\oauthor{\bsnm{Cahill}, \binits{R.A.}}:
Inter-user variation in the interpretation of near infrared perfusion imaging
  using indocyanine green in colorectal surgery.
Surgical Endoscopy,
1--8
(2021)
\end{botherref}
\endbibitem

\bibitem{Maier-Hein2022}
\begin{barticle}
\bauthor{\bsnm{Maier-Hein}, \binits{L.}},
\bauthor{\bsnm{Eisenmann}, \binits{M.}},
\bauthor{\bsnm{Sarikaya}, \binits{D.}},
\bauthor{\bsnm{M{\"{a}}rz}, \binits{K.}},
\bauthor{\bsnm{Collins}, \binits{T.}},
\bauthor{\bsnm{Malpani}, \binits{A.}},
\bauthor{\bsnm{Fallert}, \binits{J.}},
\bauthor{\bsnm{Feussner}, \binits{H.}},
\bauthor{\bsnm{Giannarou}, \binits{S.}},
\bauthor{\bsnm{Mascagni}, \binits{P.}},
\bauthor{\bsnm{Nakawala}, \binits{H.}},
\bauthor{\bsnm{Park}, \binits{A.}},
\bauthor{\bsnm{Pugh}, \binits{C.}},
\bauthor{\bsnm{Stoyanov}, \binits{D.}},
\bauthor{\bsnm{Vedula}, \binits{S.S.}},
\bauthor{\bsnm{Cleary}, \binits{K.}},
\bauthor{\bsnm{Fichtinger}, \binits{G.}},
\bauthor{\bsnm{Forestier}, \binits{G.}},
\bauthor{\bsnm{Gibaud}, \binits{B.}},
\bauthor{\bsnm{Grantcharov}, \binits{T.}},
\bauthor{\bsnm{Hashizume}, \binits{M.}},
\bauthor{\bsnm{Heckmann-N{\"{o}}tzel}, \binits{D.}},
\bauthor{\bsnm{Kenngott}, \binits{H.G.}},
\bauthor{\bsnm{Kikinis}, \binits{R.}},
\bauthor{\bsnm{M{\"{u}}ndermann}, \binits{L.}},
\bauthor{\bsnm{Navab}, \binits{N.}},
\bauthor{\bsnm{Onogur}, \binits{S.}},
\bauthor{\bsnm{Ro{\ss}}, \binits{T.}},
\bauthor{\bsnm{Sznitman}, \binits{R.}},
\bauthor{\bsnm{Taylor}, \binits{R.H.}},
\bauthor{\bsnm{Tizabi}, \binits{M.D.}},
\bauthor{\bsnm{Wagner}, \binits{M.}},
\bauthor{\bsnm{Hager}, \binits{G.D.}},
\bauthor{\bsnm{Neumuth}, \binits{T.}},
\bauthor{\bsnm{Padoy}, \binits{N.}},
\bauthor{\bsnm{Collins}, \binits{J.}},
\bauthor{\bsnm{Gockel}, \binits{I.}},
\bauthor{\bsnm{Goedeke}, \binits{J.}},
\bauthor{\bsnm{Hashimoto}, \binits{D.A.}},
\bauthor{\bsnm{Joyeux}, \binits{L.}},
\bauthor{\bsnm{Lam}, \binits{K.}},
\bauthor{\bsnm{Leff}, \binits{D.R.}},
\bauthor{\bsnm{Madani}, \binits{A.}},
\bauthor{\bsnm{Marcus}, \binits{H.J.}},
\bauthor{\bsnm{Meireles}, \binits{O.}},
\bauthor{\bsnm{Seitel}, \binits{A.}},
\bauthor{\bsnm{Teber}, \binits{D.}},
\bauthor{\bsnm{{\"{U}}ckert}, \binits{F.}},
\bauthor{\bsnm{M{\"{u}}ller-Stich}, \binits{B.P.}},
\bauthor{\bsnm{Jannin}, \binits{P.}},
\bauthor{\bsnm{Speidel}, \binits{S.}}:
\batitle{{Surgical data science – from concepts toward clinical
  translation}}.
\bjtitle{Medical Image Analysis}
\bvolume{76},
\bfpage{102306}
(\byear{2022})
{\href{https://arxiv.org/abs/2011.02284}{{arXiv:2011.02284}}}.
\doiurl{10.1016/j.media.2021.102306}
\end{barticle}
\endbibitem

\bibitem{soares2022multisensor}
\begin{botherref}
\oauthor{\bsnm{Soares}, \binits{A.S.}},
\oauthor{\bsnm{Bano}, \binits{S.}},
\oauthor{\bsnm{Clancy}, \binits{N.T.}},
\oauthor{\bsnm{Stoyanov}, \binits{D.}},
\oauthor{\bsnm{Lovat}, \binits{L.B.}},
\oauthor{\bsnm{Chand}, \binits{M.}}:
Multisensor perfusion assessment cohort study: Preliminary evidence toward a
  standardized assessment of indocyanine green fluorescence in colorectal
  surgery.
Surgery
(2022)
\end{botherref}
\endbibitem

\bibitem{TRIPOD}
\begin{botherref}
Transparent reporting of a multivariable prediction model for individual
  prognosis or diagnosis (tripod): The tripod statement.
Annals of Internal Medicine
\textbf{162}(1),
55--63
(2015)
{\href{https://arxiv.org/abs/https://doi.org/10.7326/M14-0697}{{https://doi.org/10.7326/M14-0697}}}.
\doiurl{10.7326/M14-0697}.
PMID: 25560714
\end{botherref}
\endbibitem

\bibitem{howard2018fastai}
\begin{botherref}
\oauthor{\bsnm{Howard}, \binits{J.}}, et al.:
The fast.ai deep learning library.
GitHub
(2018)
\end{botherref}
\endbibitem

\bibitem{Schneider2012}
\begin{botherref}
\oauthor{\bsnm{Schneider}, \binits{C.A.}},
\oauthor{\bsnm{Rasband}, \binits{W.S.}},
\oauthor{\bsnm{Eliceiri}, \binits{K.W.}}:
{NIH Image to ImageJ: 25 years of image analysis}
(2012).
\doiurl{10.1038/nmeth.2089}.
\url{http://www.nature.com/articles/nmeth.2089}
\end{botherref}
\endbibitem

\bibitem{wada2017icg}
\begin{barticle}
\bauthor{\bsnm{Wada}, \binits{T.}},
\bauthor{\bsnm{Kawada}, \binits{K.}},
\bauthor{\bsnm{Takahashi}, \binits{R.}},
\bauthor{\bsnm{Yoshitomi}, \binits{M.}},
\bauthor{\bsnm{Hida}, \binits{K.}},
\bauthor{\bsnm{Hasegawa}, \binits{S.}},
\bauthor{\bsnm{Sakai}, \binits{Y.}}:
\batitle{Icg fluorescence imaging for quantitative evaluation of colonic
  perfusion in laparoscopic colorectal surgery}.
\bjtitle{Surgical endoscopy}
\bvolume{31}(\bissue{10}),
\bfpage{4184}--\blpage{4193}
(\byear{2017})
\end{barticle}
\endbibitem

\bibitem{son2019quantitative}
\begin{barticle}
\bauthor{\bsnm{Son}, \binits{G.M.}},
\bauthor{\bsnm{Kwon}, \binits{M.S.}},
\bauthor{\bsnm{Kim}, \binits{Y.}},
\bauthor{\bsnm{Kim}, \binits{J.}},
\bauthor{\bsnm{Kim}, \binits{S.H.}},
\bauthor{\bsnm{Lee}, \binits{J.W.}}:
\batitle{Quantitative analysis of colon perfusion pattern using indocyanine
  green (icg) angiography in laparoscopic colorectal surgery}.
\bjtitle{Surgical endoscopy}
\bvolume{33}(\bissue{5}),
\bfpage{1640}--\blpage{1649}
(\byear{2019})
\end{barticle}
\endbibitem

\end{thebibliography}


\end{document}